# Nanoinstabilities of $Cu_2O$ porous nanostructured films as driven by nanocurvature effect and thermal activation effect

Yiqi Zhu[1], Ji Ma[2], Jiangbin Su[1,2,3*], Lei Zhou[2], Meiping Jiang[2*], Xianfang Zhu[3]

1 School of Electrical & Information Engineering, Jiangsu University of Technology, Changzhou 213001, China

2 School of Mathematics & Physics, Changzhou University, Changzhou 213164, China

3 China-Australia Joint Laboratory for Functional Nanomaterials and Department of Physics, Xiamen University, Xiamen 361005, China

* Corresponding authors. Email: jbsu@cczu.edu.cn (J. B. Su); mpjiang@cczu.edu.cn (M. P. Jiang)

## Abstract

In this work, the instabilities at the nanoscale (i.e. nanoinstabilities) of triangular pyramids-like $Cu_2O$ porous nanostructured films (PNFs) are studied by heating treatments under different atmosphere and temperature. It is found that the nanoscale building triangular-pyramids turn round preferentially at the sharp angles and/or coalesce with their contacting ones by directional diffusion and plastic flow of atoms, which are driven by the nonuniformly-distributed surface nanocurvature. As a result, the triangular pyramids become quasi-sphere shape and the PNF evolves into a big, dense particles film. It is also observed that the heating or thermal activation effect efficiently promotes the reduction or oxidation of $Cu_2O$ pyramids and the crystallization or growth of the as-achieved Cu or CuO grains. The above physical and chemical instabilities or changes at the nanoscale of $Cu_2O$ PNFs can be well accounted for by the combined mechanism of nanocurvature effect and thermal activation effect. The nanocurvature effect can lower the energy barrier for the atom diffusion or plastic flow and lower the activation energy for the chemical reactions, while the thermal activation effect can supply the required kinetic energy or activation energy and make the atomic transportations and reactions kinetically possible. The findings reveal the evolution laws of morphology, crystal structure and composition of triangular pyramids-like $Cu_2O$ PNF during heating treatments, which can further be extended to other types of $Cu_2O$ PNFs. Also, the findings have important implications for the nanoinstabilities of $Cu_2O$

PNFs-based devices, especially those working at a high temperature.

**Keywords:** $Cu_2O$; porous nanostructured films; nanoinstability; nanocurvature effect; thermal activation effect.

## 1. Introduction

Cuprous oxide ($Cu_2O$) is an important p-type transition metal oxide semiconductor material. It has the advantages of low-cost, non-toxicity and abundant copper sources and can be potentially applied in many fields such as gas sensors [1,2], solar cells [3,4] and photocatalysts [5-7]. In order to improve the performance of the above $Cu_2O$-based surface-sensitive devices and materials, the researchers tend to prepare $Cu_2O$ thin films of porous structure [2,6]. Among them, there is a special kind of porous films which is composed by plenty of solid and/or hollow nanostructures (e.g., nanoparticles, nanoligaments, nanowires, nanoplates, nanocavities, nanopores or nanochannels) and exhibits a distinctive structure of porous nanostructured films or porous nanostructure-films (PNFs for short). Due to the unique PNF structure along with the increased film porosity and surface area, it is expected that such PNFs may have improved or enhanced performance in contrast to the traditional porous thin films and solid thin films [8,9]. However, since the building nanostructures are highly curved and limited in space within nanoscale, the surface energy is dramatically increased [10,11] and will cause an intrinsic structural or physical instability in the PNFs. This phenomena are the so-called (surface) nanocurvature effect in a broad sense [12,13]. As a consequence, the PNFs tend to change the morphology and crystal structure of their building nanostructures thermodynamically. In addition, the valence state of Cu element in $Cu_2O$ is +1 ($Cu^{1+}$), which is also unstable as driven by the tendency of decreasing the free energy [14]. Thus, the $Cu^+$ in $Cu_2O$ seems easy to be oxidized into $Cu^{2+}$ or reduced into $Cu^0$ especially at an elevated temperature or under a certain heating treatment. We call this thermal activation-induced redox a chemical instability in the broad sense. Also note that the nanocurvature effect may influence the chemical instability of $Cu_2O$ of PNF structures to a certain extent. Based on the above analysis, it can be inferred that both the physical and chemical instabilities of $Cu_2O$ PNFs have negative effects on the performance, stability and lifetime of the $Cu_2O$ PNFs-based devices and materials. Therefore, the study on the physical and chemical instabilities of $Cu_2O$ PNFs especially the effects of nanocurvature and thermal activation is crucial and imperative. In the existing literature, the study on the instability of $Cu_2O$ mainly concentrates on two aspects: (1) in the process of photoelectrochemistry or photocatalysis, the stability difference between crystallographic planes caused by their different energy and adsorption of ions in the solution [15-21]; (2) in the process of heating treatment, the morphology and crystal structure changes due to the redox or grain growth [22-24]. Nevertheless, most of these studies aim at micro-sized particles [15,17-19,24] or dense films [16,21,23], whose nanocurvature effect is not obvious and also has not been studied. Even in the little study of the instability of porous $Cu_2O$ [20] and nano-sized $Cu_2O$ [22], the specific nanocurvature effect and the key influence of thermal activation on the instability of $Cu_2O$ have not been paid sufficient attention.

With the above considerations, in this paper we particularly study the physical and chemical instabilities at the nanoscale of triangular pyramids-like $Cu_2O$ PNFs by heating treatments under different atmosphere and temperature. It is found that at a low temperature of 200 ℃, the building triangular pyramids of PNF turn round preferentially at the sharp angles and/or coalesce with their contacting ones. As a result, the triangular pyramids become quasi-sphere shape and the PNF evolves into a big, dense particles film. When heated at a higher temperature, the $Cu_2O$ is reduced or oxidized from the pyramid surface, and the heating can efficiently promote the redox of $Cu_2O$ pyramids and the crystallization or growth of the as-formed Cu or CuO grains. Based on the observations and an in-depth analysis, a new mechanism combined nanocurvature effect with thermal activation effect is proposed to elucidate the observed nanophenomena.

## 2. Experimental section

$Cu_2O$ thin films of a triangular pyramids-like PNF structure were fabricated by a radio-frequency balanced magnetron sputtering approach at a substrate bias voltage of +50 V. The details of the fabrication can be found in Ref. [25]. In this bias deposition, the tip charging effect [25,26] was demonstrated to dominate the formation process of the PNFs. To investigate the physical and chemical instabilities of the as-prepared $Cu_2O$ PNFs, several kinds of heating treatments such as rapid thermal processing (RTP), high vacuum annealing (HVA), dry-oxygen oxidation and wet-oxygen oxidation were respectively carried out at 200−500 ℃ for a duration of 30 min. In order to further explore the evolution details of $Cu_2O$ PNFs, we also stopped the heating at an intermediate time such as 15 min. During the RTP, a nitrogen gas of 99.999 wt.% high purity was continuously introduced to the quartz chamber; for the HVA, a base pressure of chamber of $6\times10^{-4}$ Pa was achieved prior to the annealing;

for the dry- and wet-oxygen oxidation, a high purity oxygen gas (99.999 wt.%) without and with water vapor (obtained by heating deionized water to 80 ℃) was respectively introduced to the quartz tube of oxidation furnace. The as-annealed samples were then characterized by a powder X-ray diffractometer (XRD, RIGAKU D/Max 2500 PC) and a field-emission scanning electron microscope (SEM, ZEISS SUPRA 55) to study their changes in crystal structure, chemical composition and surface morphology.

## 3. Results and discussion

Fig. 1 shows the XRD result of the PNFs as-deposited and after RTP at 200 ℃ for different time. It can be observed that all the samples before and after the RTP exhibit a preferential orientation of $Cu_2O$ (111) (JCPDS no. 65-3288) and no diffraction peaks of other substances such as Cu and CuO appear after the RTP. Furthermore, the intensity of diffraction peak (111) seems to change very little. It demonstrates that during the RTP at a relatively low temperature of 200 ℃ the $Cu_2O$ PNF tends to keep its chemical composition and crystal structure. However, the SEM images in Fig. 2 show a notable morphology evolution in the $Cu_2O$ PNF when RTP at 200 °C with the increase of annealing time. As shown in Fig. 2(a), it is found that before the RTP the $Cu_2O$ PNF exhibits a unique triangular pyramids-like PNF structure. The nanoscale triangular pyramids with cut sharp edges and corners (or tips) appear on the PNF surface and little gluing can be seen between each other. The average side length of the pyramids is ~48 nm and the average pore diameter is ~13 nm. With the RTP turned on, as shown in Fig. 2(b-c), the edges and corners of the triangular pyramids become round gradually and the neighbouring triangular pyramids contact and coalesce together. In details, after RTP for 15 min (an intermediate stage for example), the triangular pyramids partially coalesce and the PNF presents a wrinkly (peanut-like or necked) and relatively dense film surface along with some tiny cracks (see Fig. 2(b)). After RTP for 30 min, the building triangular pyramids of the PNF coalesce further and finally turn into some big particles with an average diameter of ~72 nm (see Fig. 2(c)). The fast reshaping and coalescence demonstrate an intriguing atom diffusion and plastic flow or wetting effect [12] especially at the contact locations of triangular pyramids with a fast, massive atom filling-in. In this way, the contact locations gather more and more atoms, resulting in the local coalescence or welding at the contact locations. Meanwhile, the cracks become less and wider remarkably and the PNF gets denser, which can be attributed to the coalescence of triangular pyramids and the filling of material atoms in the cracks.

The above findings demonstrate a preferential evolution in surface morphology over that in chemical composition and crystal structure. It implies that there is a relatively lower energy barrier for the reshaping and coalescence to that for the crystallization or grain growth and the chemical reactions. Such a preferential instability in surface morphology or structural instability as induced by heating treatment can be fully accounted for by our proposed nanocurvature effect and the thermal activation effect. For the nanocurvature effect of an isolated triangular pyramid, as illustrated in Fig. 3(a), the surface is highly curved at the sharp convex angles (e.g., Location 1) and the curvature approaches to positive infinity (i.e. $\rho\rightarrow+\infty$). Such an extremely high positive nanocurvature will cause an additional tensile stress [12,13] on the electron cloud structure of surface atoms at the sharp angles. Therefore, the vibration frequency of surface atoms will be decreased and the "Debye temperature" [12,13] or melting point will be lowered down and cause the sharp angles to melt and the atoms therein to migrate or even flow. At the bottom or side (e.g., Location 2), the surface of the triangular pyramid is flat without any curvature (i.e. $\rho\rightarrow 0$). It can supply the lower-energy location for the aggregation of atoms coming from the energetic sharp angles. When two triangular pyramids get close and contact together, as illustrated in Fig. 3(b), a negative nanocurvature which approaches to negative infinity (i.e. $\rho\rightarrow-\infty$) will form at the contact location (i.e. the sharp concave angles, Location 3 for example). In contrast to the positive nanocurvature, the negative nanocurvature will cause an additional compressive stress [12,13] on the electron cloud structure of surface atoms. This compressive stress will lead to a speeding up of the vibration of surface atoms and thus increase the "Debye temperature" or melting point and induce the sharp concave angles to condense and capture other atoms. Nevertheless, both the positive and negative nanocurvatures will dramatically increase the surface energy and lower the energy barriers to be $\Delta G^*$ (see a→b in Fig. 4) for the out-going or in-coming diffusion of surface atoms and even the plastic flow of massive atoms. As a result, both the isolated and the contacted triangular pyramids at the nanoscale would exhibit an intrinsic structural instability thermodynamically.

Although the nonuniformly-distributed nanocurvature on the triangular pyramid can cause the above structural instability tendency thermodynamically, a further assistance from external excitation such as energetic electron or ion beam irradiation [10-12] or heating is still needed to realize kinetically the corresponding mass transportation and morphology change. In the present case of RTP, the heating can intensify the thermal movement of atoms and increase their kinetic energy. This increased kinetic energy can impel the surface atoms to pass over the suppressed energy barrier (see b→c in Fig. 4) and realize the atom transportations such as surface diffusion and plastic flow. In doing so, the heating can cause the diffusion of surface atoms or plastic flow of massive atoms and finally make the reshaping and coalescence of triangular pyramids kinetically possible. This phenomena is the as-called thermal activation effect in a broad sense.

In the following, we present the processes of reshaping and coalescence of the triangular pyramids. As shown in Fig. 4, when the triangular pyramid is subjected to RTP, the surface atoms will get enough kinetic energy to pass over the suppressed energy barrier caused by the nanocurvature effect. On the other hand, as shown in Fig. 3, the sharp convex angles or tips behave like a wetting effect and the energetic atoms herein will diffuse or flow to the bottom or side along the pyramid surface as driven by the nanocurvature effect. When two or more energetic tips of triangular pyramids contact together, they will coalesce together instantly and the surface atoms at the bottom or side or in-coming from the convex angles will (further) diffuse or flow to the sharp concave angles as-formed at the contact location. With the atom diffusion and plastic flow going on, the sharp convex angles turn rounded and the sharp concave angles are quickly filled by the atom adjustment or rearrangement. In this way, the isolated triangular pyramids tend to form spherical particles (see Fig. 3(a)), and the contacted triangular pyramids will coalesce into a necked peanut-like configuration at first and then also turn into spherical particles (see Fig. 3(b)). The surface nanocurvature of as-formed spherical particles tends to be uniform ($\rho=2/r$), in which $r$ is the radius of particles. As demonstrated by the SEM images in Fig. 2, the triangular pyramids-like PNF evolved into a layer of quasi-spherical particles after 30 min RTP. We also note that the as-formed particles will further contact and coalesce with their adjacent particles during the heating. As a result, the particles are contacting with each other closely in the annealed film. We believe that as long as the duration is sufficient, a full coalescence of all the contact particles will be achieved and the ideal separated spherical particles with a minimum surface area can also be observed.

In the above, we have demonstrated the crucial effects of nanocurvature and thermal activation on the morphology changes (physical instability) of $Cu_2O$ PNFs. In fact, these two effects will also affect the changes in crystal structure (another kind of physical instability) and chemical composition (chemical instability) of $Cu_2O$ PNFs. In the following, we in particular study the influence of thermal activation on the changes in crystal structure and chemical composition.

In Fig. 5, it shows the XRD results of the $Cu_2O$ PNFs after 30 min-RTP and 30 min-HVA at different temperature. At 200 ℃, as shown in Fig. 5(a,b), the annealed films both exhibit a pure composition of $Cu_2O$ with a preferential orientation of (111), which is nearly the same as that in the as-deposited $Cu_2O$ film. With the increase of annealing temperature, as shown in Fig. 5(a-c), both of the diffraction peaks of $Cu_2O$ (111) obtained by RTP and HVA first intensify, then weaken, or even disappear completely. At the same time, the peaks of Cu (111) (JCPDS no. 65-9026) appear and intensify continuously in both the annealed films. By applying the Scherrer Equation, we obtained that the average sizes of $Cu_2O$ (111) grains and Cu (111) grains both increase fast at first and then slow down (for $Cu_2O$ grains, they will disappear finally, see Fig. 5(d)). It demonstrates that during the RTP and HVA, both the $Cu_2O$ are being reduced to Cu continuously, and the as-formed Cu grains and the residual $Cu_2O$ grains in the films are growing up. In other words, the heating or thermal activation effect can effectively promote the reduction of $Cu_2O$ pyramids and the growth of $Cu_2O$ and Cu grains. Also note that the peak intensity of $Cu_2O$ (111) decreases when the PNFs are annealed at a high temperature, such as 400−500 ℃ for RTP and 300−500 ℃ for HVA. This phenomenon can be mainly attributed to the decreasing in quantity of $Cu_2O$ grains during the reduction of $Cu_2O$.

Fig. 6 further shows the XRD results of the $Cu_2O$ PNFs annealed in dry oxygen ($O_2$) and wet oxygen ($O_2$+$H_2O$) at different temperature for a duration of 30 min. As shown in Fig. 6(a,b), the $Cu_2O$ PNFs are quickly oxidized and change their composition at 200 ℃ especially in the case of wet oxygen. When annealed at 300 ℃, both of the $Cu_2O$ in dry oxygen and wet oxygen are fully transformed into CuO (JCPDS no. 45-0937) after 30 min oxidation. It indicates that the heating or thermal activation effect can efficiently promote the oxidation of $Cu_2O$ pyramids. In (c,d) of Fig. 6, it further gives the diffraction peak intensity and grain size of CuO (002) and (111) formed by dry- and wet-oxygen oxidation against the oxidation temperature. We can observe that the intensity and the grain size of CuO (002) and (111) all increase with the annealing temperature no matter they are obtained by dry- or wet-oxygen oxidation. It demonstrates that the thermal activation effect can also activate the crystallization or growth of CuO grains. We should also note that there is a decline in the intensity of a certain kind of CuO grains, for example, CuO (002) during wet-oxygen oxidation at 300−500 ℃ and CuO (111) during dry-oxygen oxidation at 400−500 ℃ (see Fig. 6(c)). Meanwhile, the diffraction peak of the other kind of CuO grains seems to intensify conversely. This can be ascribed to the inter-inhibiting and competition of crystallization between two kinds of CuO grains.

Based on the above findings, we can conclude that, besides the intrinsic nanocurvature effect of nanostructures, the heating or thermal activation plays an important role in the redox process of $Cu_2O$ PNFs. As shown in Fig. 7, the heating can supply the activation energies for chemical reactions and thus kinetically realize and advance the reduction or oxidation of $Cu_2O$ pyramids. We should note that the required activation energy for the oxidation or reduction of nano-sized $Cu_2O$ pyramids especially at the sharp angles is much lower than that of bulk $Cu_2O$. This is because, the nanocurvature effect can dramatically increase the surface energy at the sharp angles of a triangular pyramid. As illustrated in Fig. 7, the energies of reactants and products on the nanocurved surface are both lifted up (X→X', Y→Y') and it will lower the required activation energy to be $E_{a,X'\to Y'}$ or $E_{a,Y'\to X'}$, which makes the chemical reactions especially at the sharp angles much easier. As a consequence, the energetic surface atoms at the sharp angles will be preferentially reduced or oxidized relative to the planar surface atoms and the internal atoms, as illustrated in Fig. 8. With the increase of annealing temperature or thermal activation, the Cu or CuO formed on the pyramid surface will be preferentially crystallized and grow into Cu or CuO grains (see Fig. 8). Meanwhile, in the center of the pyramid, the unreacted $Cu_2O$ grains will also grow up by absorbing their surrounding amorphous $Cu_2O$ or by coalescing their adjacent $Cu_2O$ grains. Similar to the reshaping or coalescence case, the crystallization or growth of the Cu-based grains is also realized *via* the atomic diffusion or rearrangement and thus can be also attributed to the nanocurvature effect and thermal activation effect (see Fig. 4). In contrast to the center of the pyramid, the nanocurvature effect on the pyramid surface can lower the energy barrier for atom diffusion and thus promote the amorphization or grain growth therein. It should be also noted that the reduction or oxidation of $Cu_2O$ on the pyramid surface will extend to the center as the increasing temperature until the redox is completed. Accompanying with the redox and grain growth, it is expected that the pyramids in PNFs may also adjust their morphology to a quasi-sphere shape by atom diffusion and plastic flow or coalesce with their surrounding pyramids, which is similar to the case in Figs. 2−3 as driven by the nanocurvature effect and thermal activation effect. In addition, since the reshaping and coalescence are prior to the crystallization or grain growth and the redox reaction, the corresponding energy barrier for reshaping and coalescence should be lower accordingly.

## 4. Conclusions

In this work, the nanoinstabilities of triangular pyramids-like $Cu_2O$ PNFs are studied by RTP, HVA, dry- and wet-oxygen oxidation at different temperature. It is found that when RTP at a low temperature of 200 ℃, the nanoscale building triangular-pyramids turn round preferentially at the sharp angles and/or coalesce with their contacting ones. As a result, the triangular pyramids become quasi-sphere shape and the PNF evolves into a big, dense particles film. It is also observed that during RTP, HVA, dry- and wet-oxygen oxidation at a higher temperature, the heating efficiently promotes the reduction or oxidation of $Cu_2O$ and the crystallization or growth of the as-achieved Cu or CuO grains. The above instabilities or changes at the nanoscale of $Cu_2O$ PNFs can be well accounted for by our proposed new concept of nanocurvature effect and the thermal activation effect. On one hand, the notable nanocurvature effect especially at the sharp angles of pyramids greatly lowers the energy barrier for atom diffusion or plastic flow and the activation energy for chemical reactions, which makes the changes in morphology, crystal structure and composition of $Cu_2O$ PNFs much easier thermodynamically. On the other hand, the heating or thermal activation effect impels the surface atoms to pass over the suppressed energy barrier or supplies the required activation energy for chemical reactions, and thus kinetically realize the reshaping, coalescence, grain growth and redox of $Cu_2O$ PNFs. Thus, the findings reveal the evolution laws of morphology, crystal structure and composition of a triangular pyramids-like $Cu_2O$ PNF during heating treatments, which can be extended to other types of $Cu_2O$ PNFs. Also, the findings have important implications for the nanoinstabilities of $Cu_2O$ PNFs-based devices, especially those working at a high temperature.

## Acknowledgements

This work was financially supported by the NSFC project under grant no. 11574255 and the Students' Extracurricular Innovation and Entrepreneurship Fundation of Changzhou University under grants no. 2017-07-C-39 and 2018-07-C-60.

## Figures

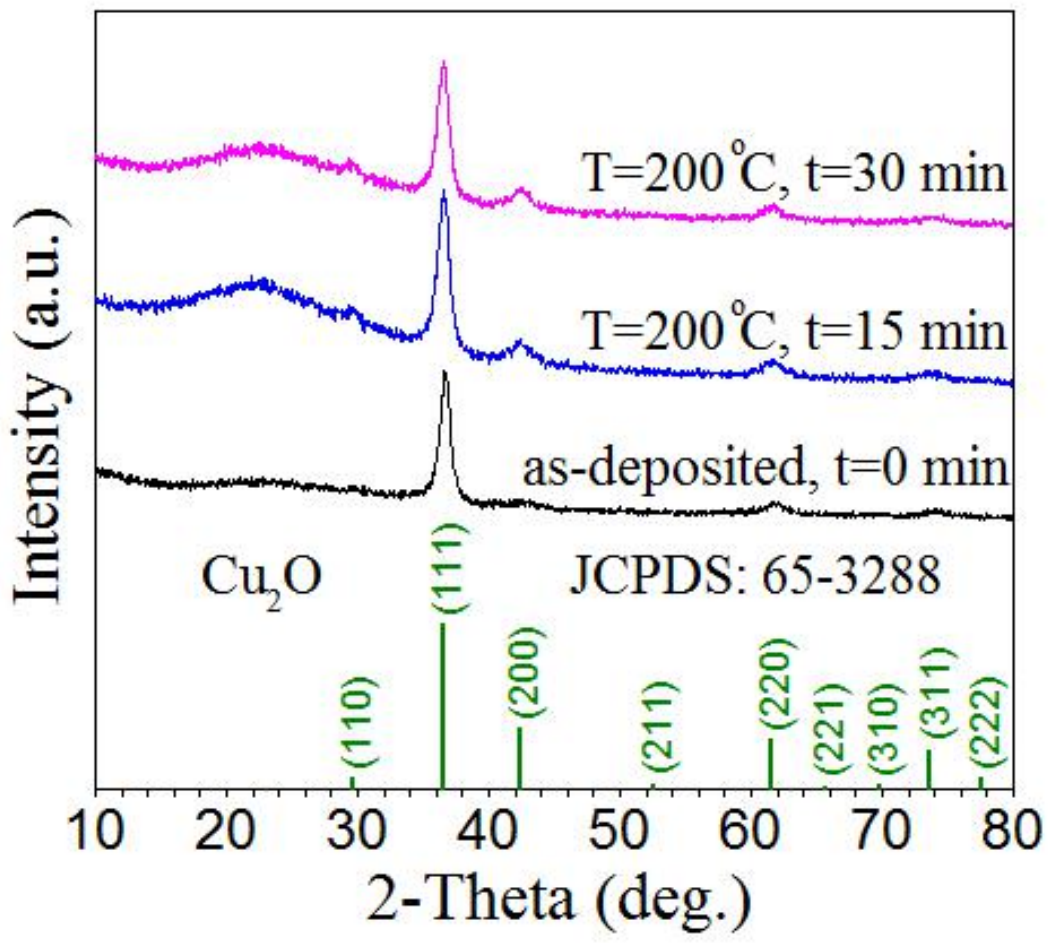

**Fig. 1** XRD patterns of the $Cu_2O$ PNF as-deposited and after RTP at 200 ˚C for different time.

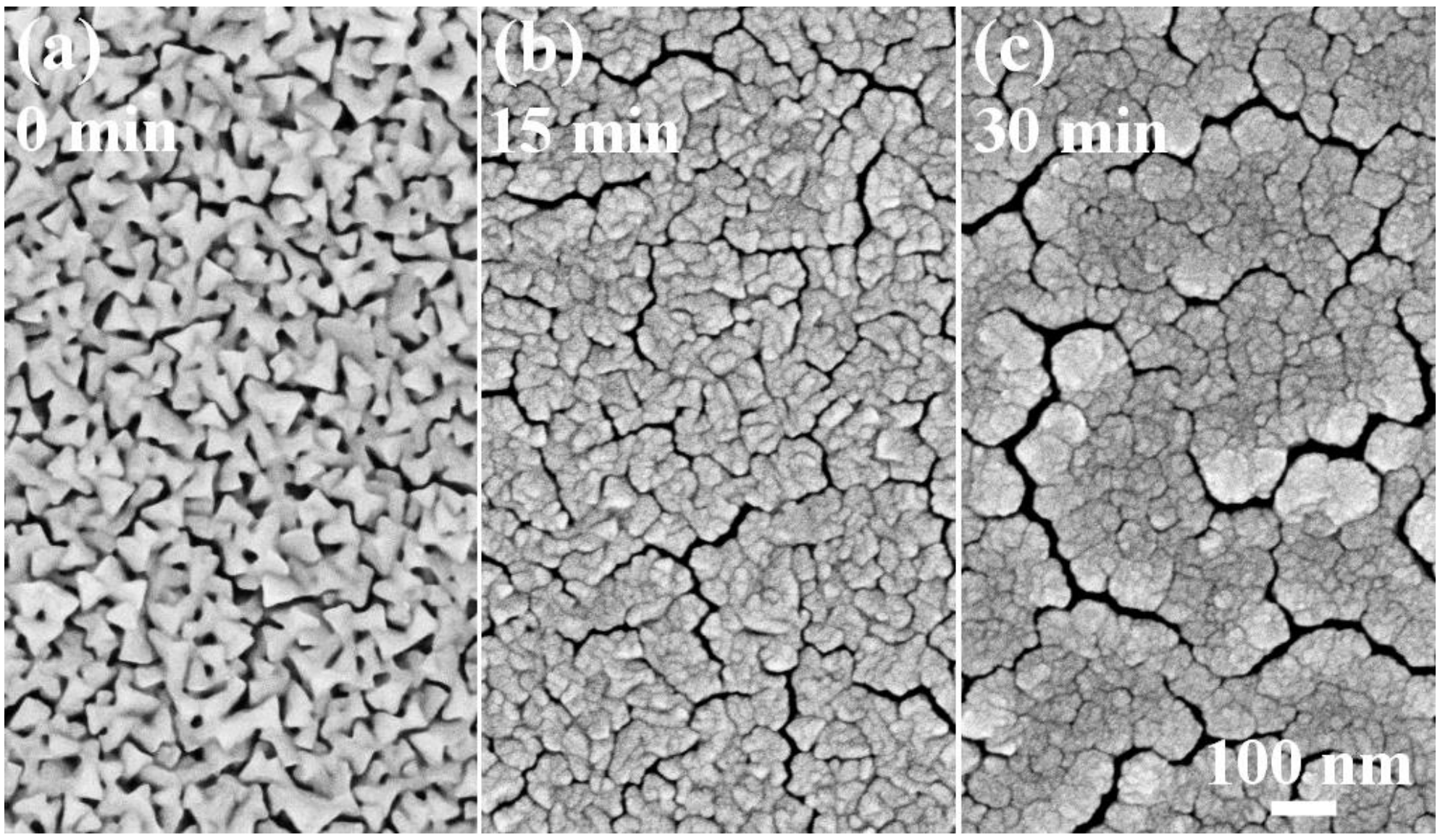

**Fig. 2** SEM images showing the morphology evolution of $Cu_2O$ PNF when RTP at 200 ˚C with the increase of annealing time: (a) t=0 min (as-deposited); (b) t=15 min; (c) t=30 min.

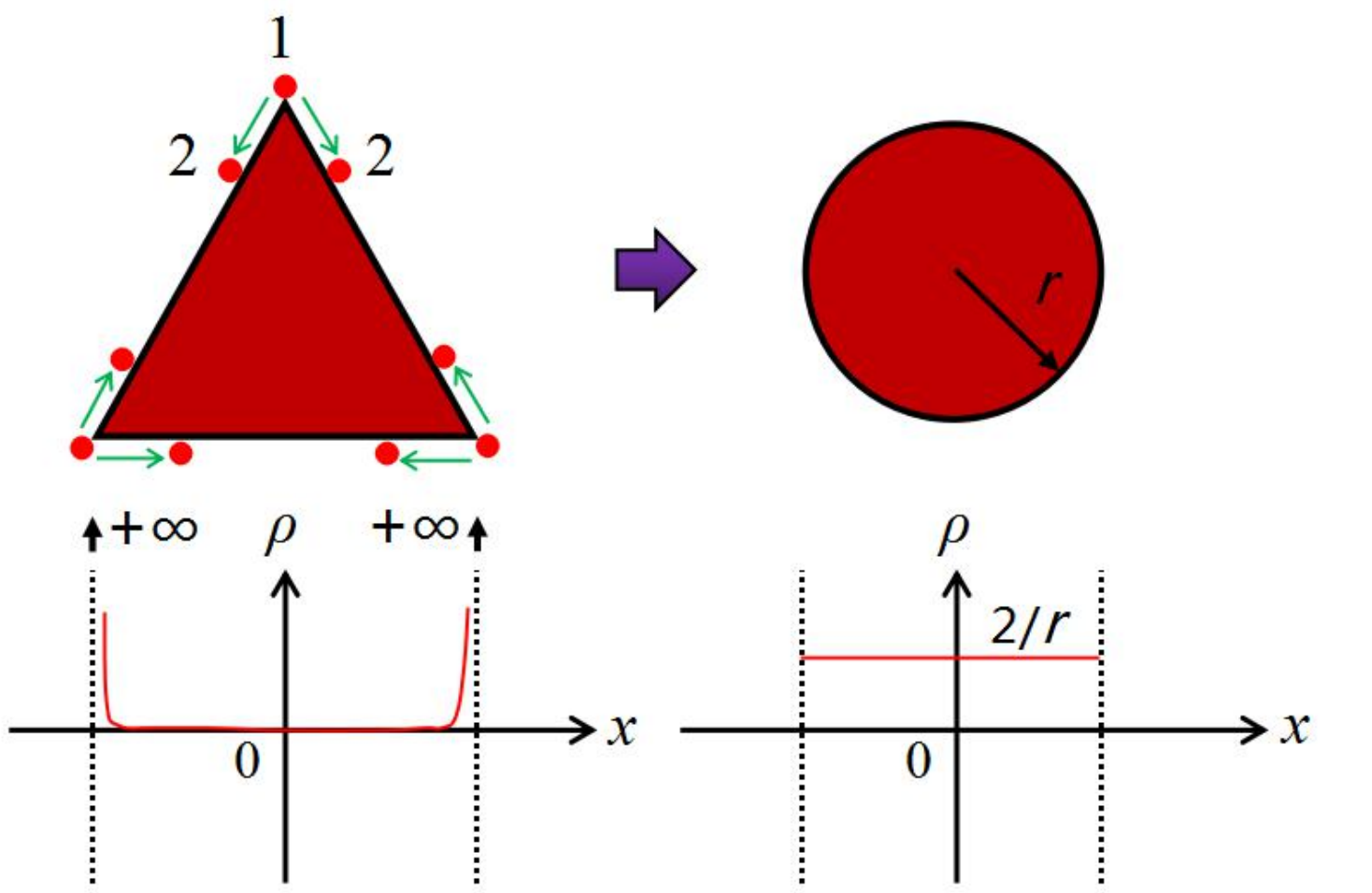

(a)

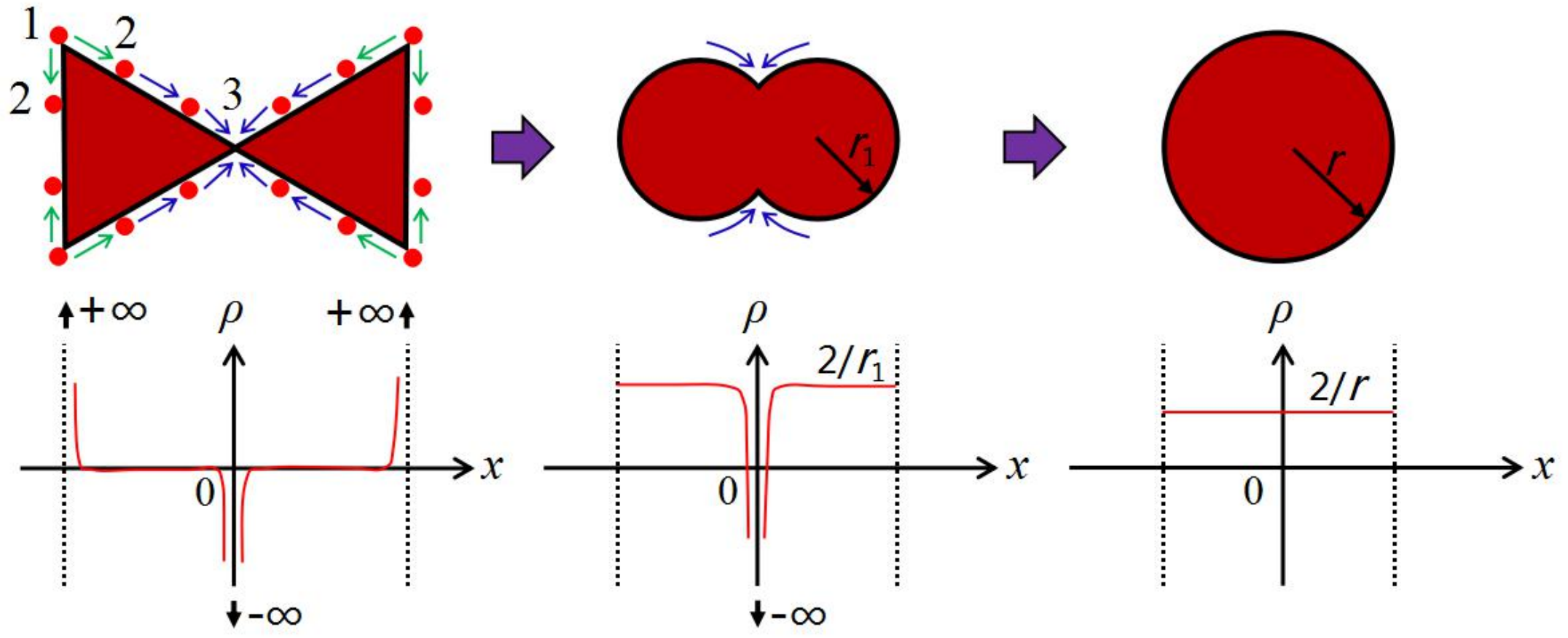

(b)

**Fig. 3** Schematic illustrations showing the directional atom diffusion (1→2, 2→3, upper figures) or plastic flow as driven by the effect of nonuniformly-distributed nanocurvature ($\rho$~$x$, lower figures) and the resulting reshaping and/or coalescence of isolated (a) and contacted (b) triangular prism(s) under heating treatments.

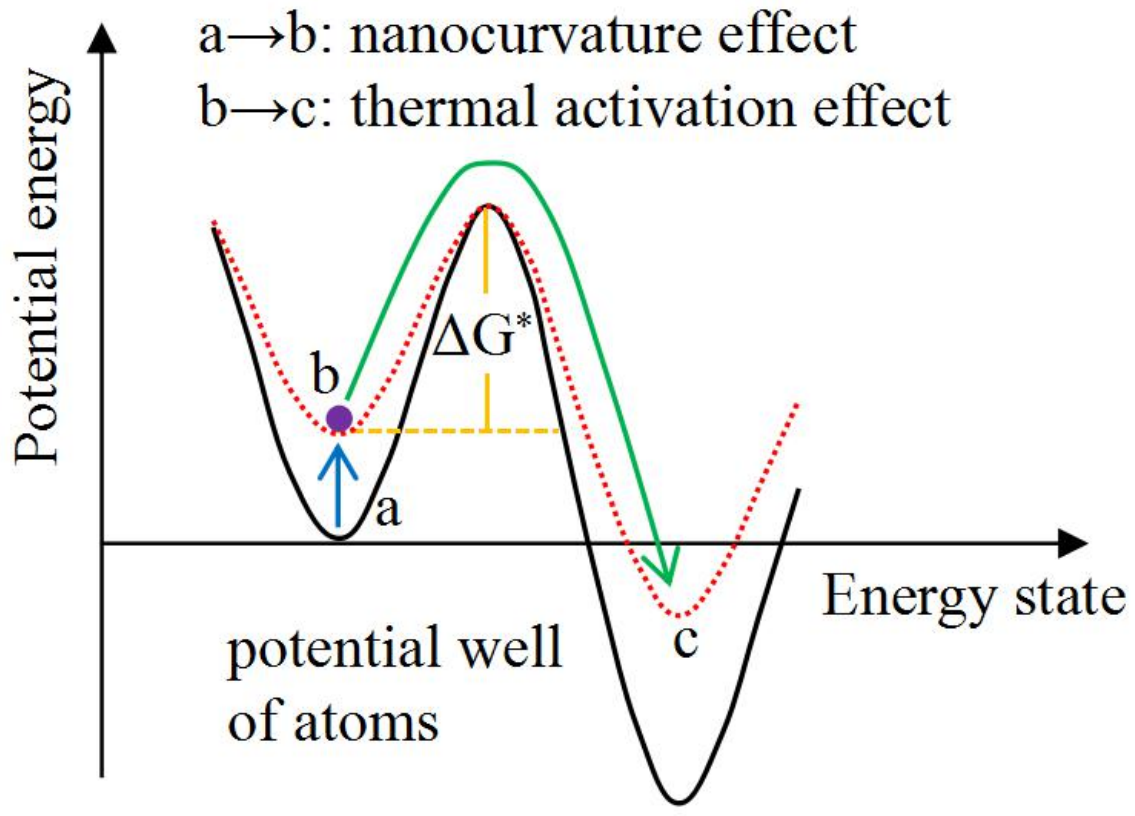

**Fig. 4** Schematic diagram showing the lifting of potential well of atoms (a→b) caused by nanocurvature effect, which causes the reduction of energy barrier $\Delta G^*$ and makes the transition of atoms from one metastable state to another (b→c) much easier.

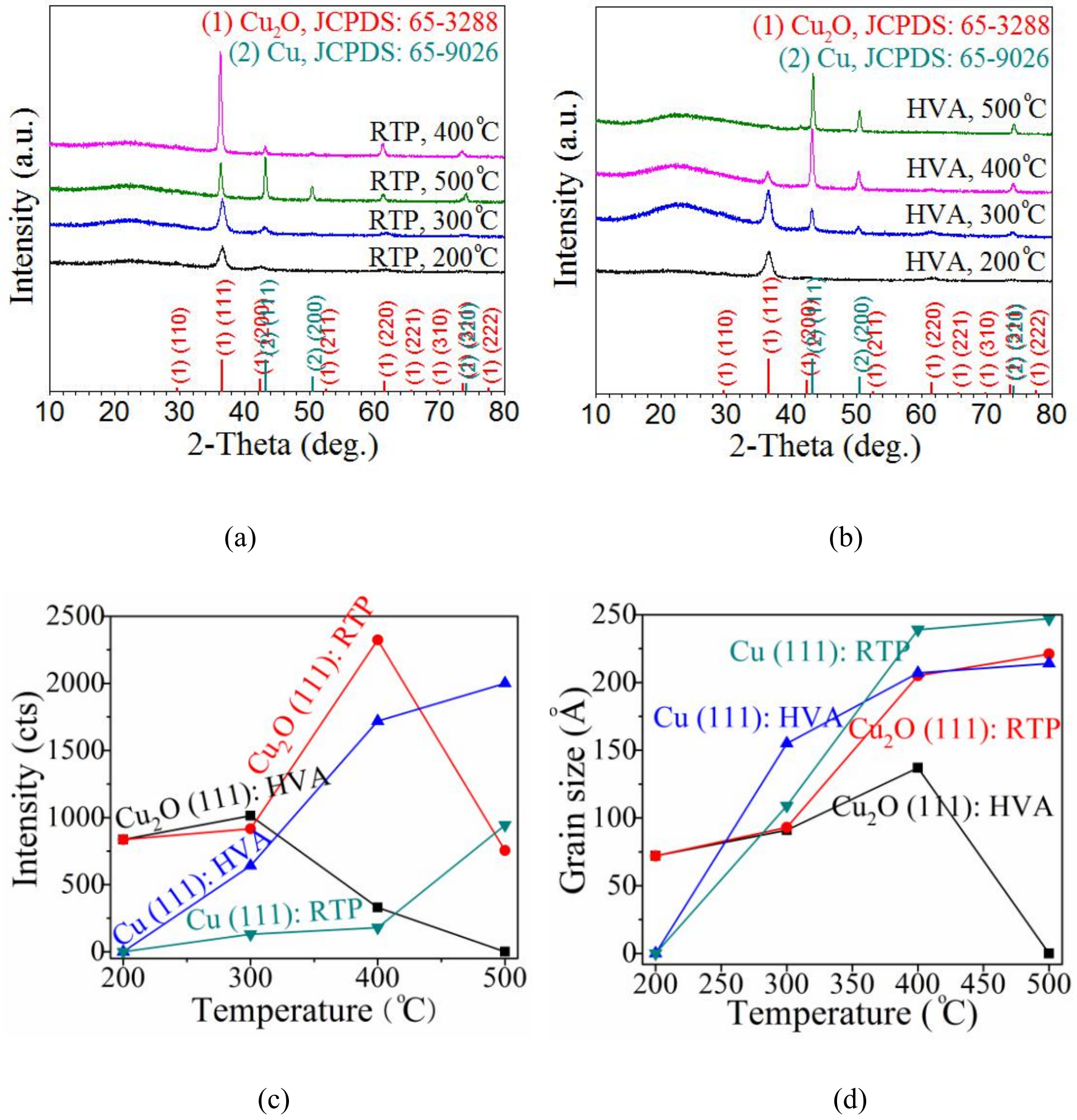

**Fig. 5** XRD patterns showing the changes in crystal structure and composition of the $Cu_2O$ PNFs after

RTP (a) and HVA (b) at different temperature; Diffraction peak intensity (c) and grain size (d) of $Cu_2O$ (111) and Cu (111) in the $Cu_2O$ PNFs after RTP and HVA at different temperature.

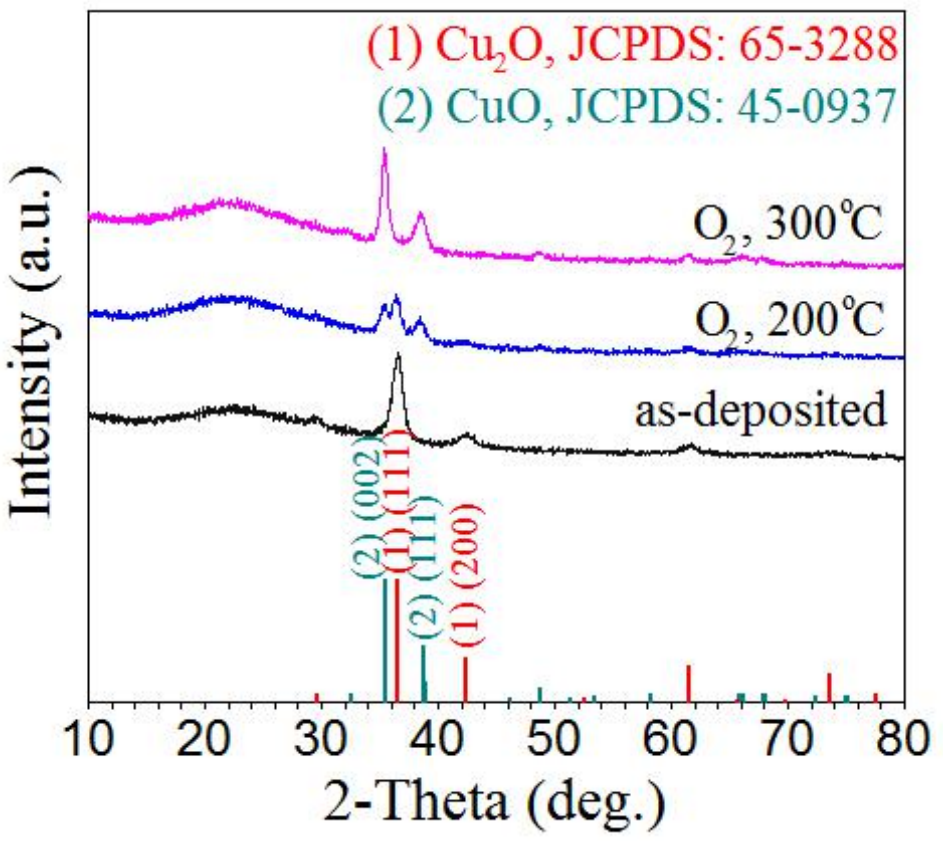

(a)

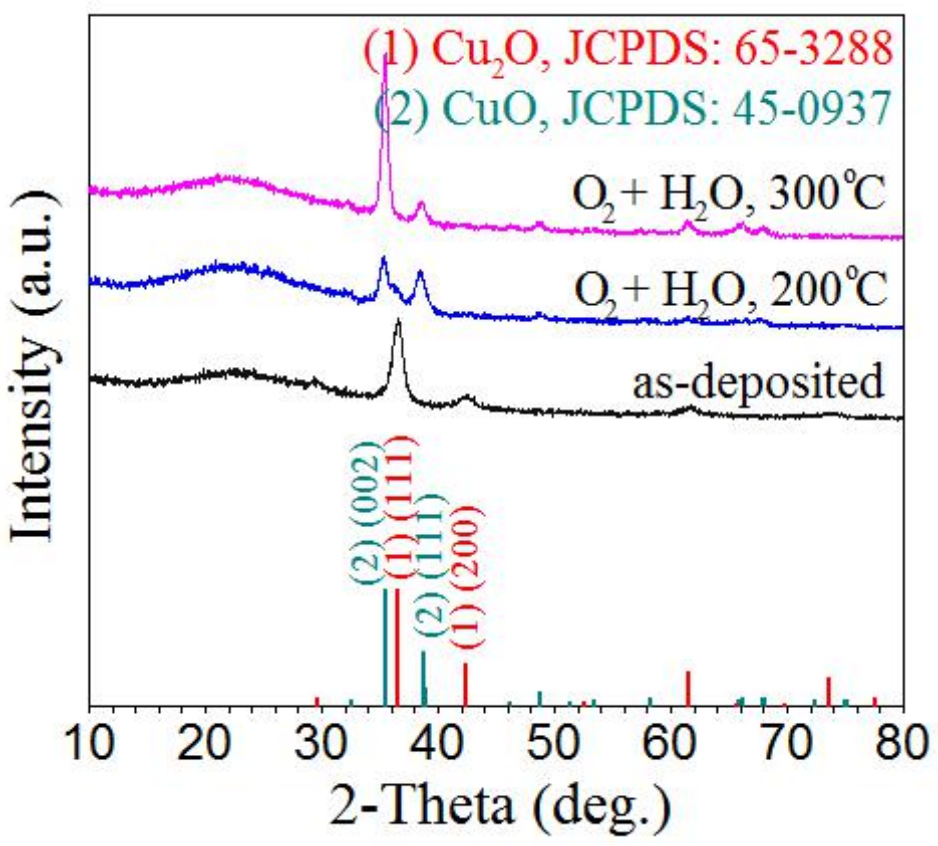

(b)

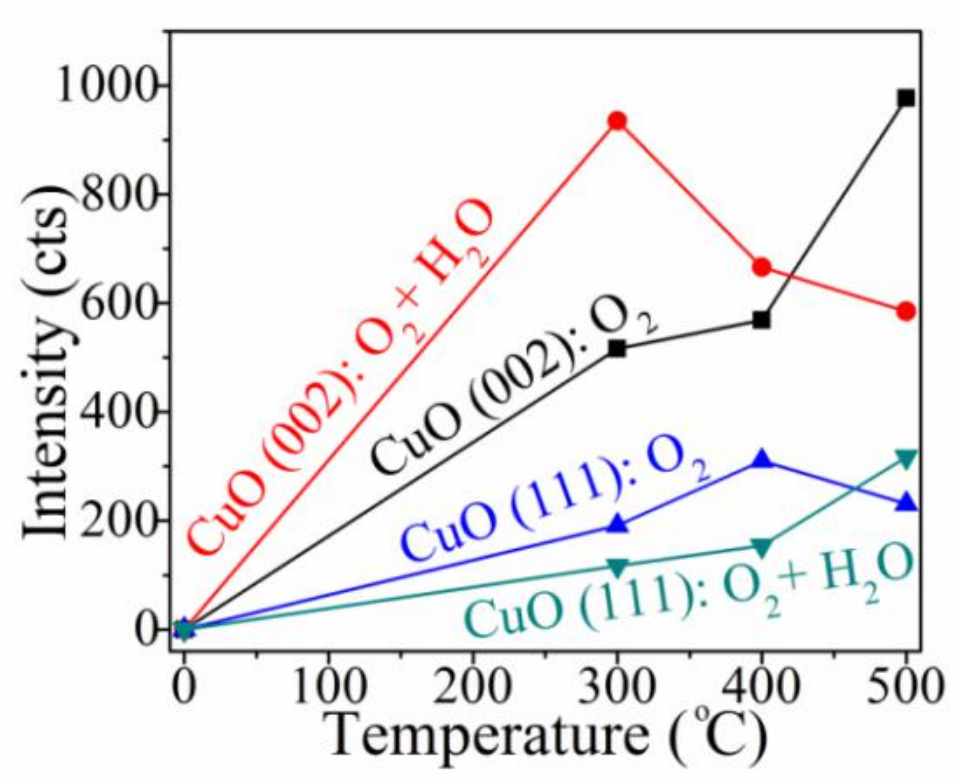

(c)

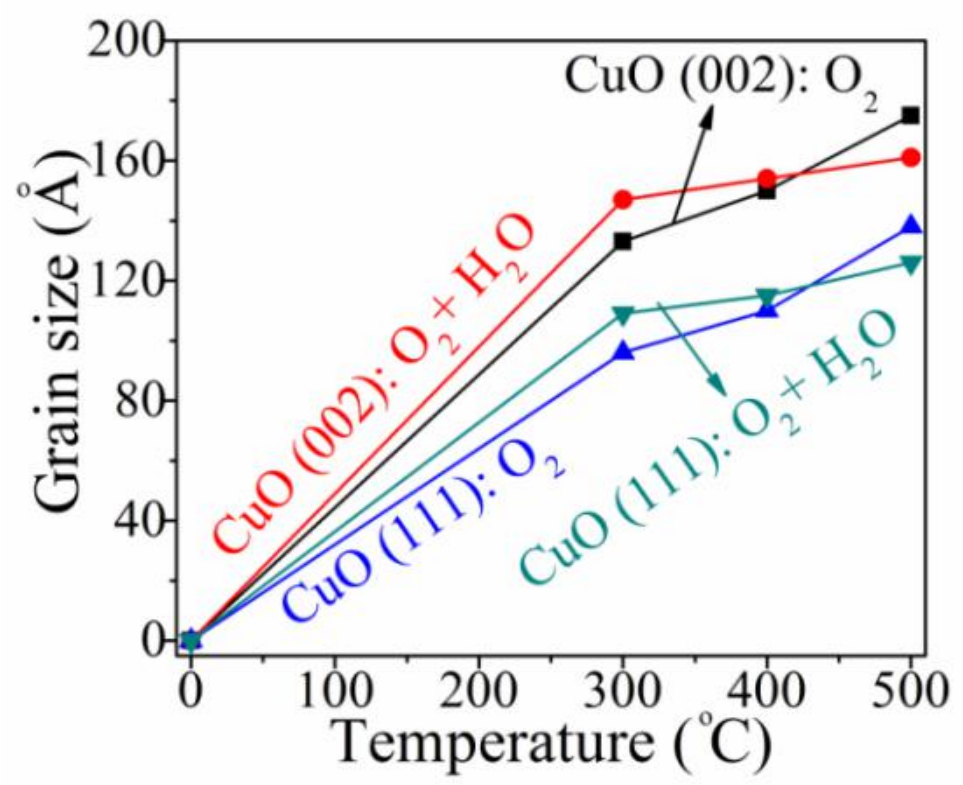

(d)

**Fig. 6** XRD patterns showing the changes in composition and crystal structure of the $Cu_2O$ PNFs annealed in dry oxygen (a) and wet oxygen (b) at different temperature; Diffraction peak intensity (c) and grain size (d) of CuO (002) and (111) formed by dry- and wet-oxygen oxidation against the oxidation temperature.

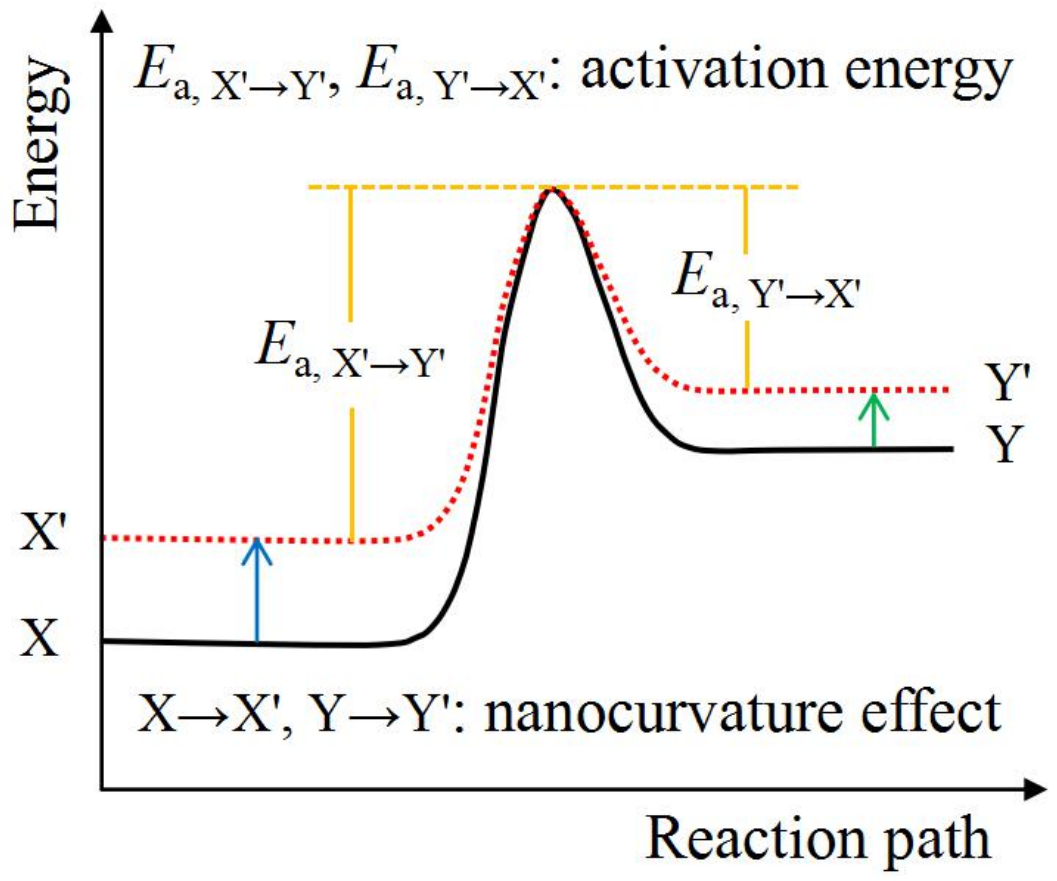

**Fig. 7** Schematic illustration showing the lifting of energies of reactants and products (X→X', Y→Y') caused by nanocurvature effect and the resulting reduction of activation energy ($E_{a,X'\rightarrow Y'}$, $E_{a,Y'\rightarrow X'}$).

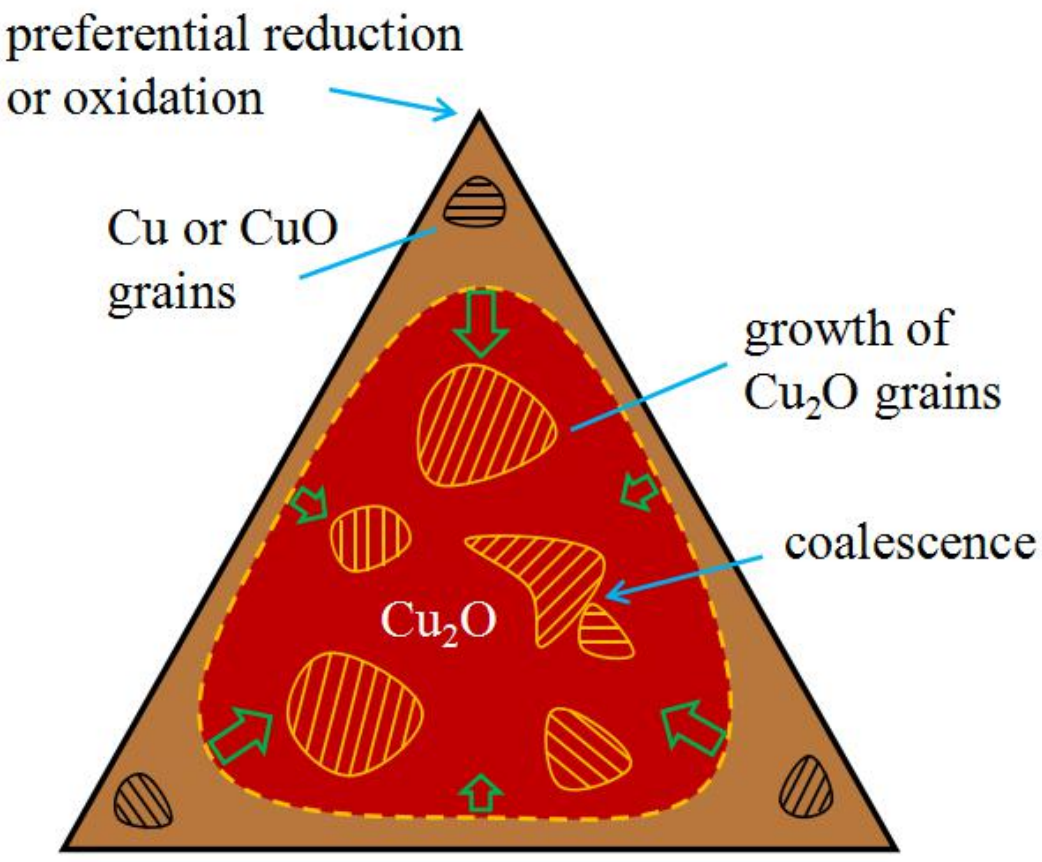

**Fig. 8** Schematic illustration showing the preferential reduction or oxidation of $Cu_2O$ triangular pyramid at the sharp angles and the growing of Cu or CuO grains.